\documentclass[reprint,prd,nofootinbib,superscriptaddress]{revtex4}
\usepackage{CJK}
\usepackage{amsfonts}
\usepackage{amsmath,slashed}
\usepackage{amssymb}
\usepackage[english]{babel}
\usepackage{graphicx}
\usepackage{epsfig}
\usepackage{bm}
\usepackage{verbatim}
\usepackage[utf8]{inputenc}
\usepackage{booktabs}
\usepackage{multirow}
\usepackage{subfig}
\usepackage{slashed}
\usepackage{xcolor}
\usepackage[colorlinks=true,urlcolor=red,citecolor=red]{hyperref}
\usepackage[font=small]{caption}
\usepackage{float}
\usepackage{blindtext}
\usepackage{placeins}

\newcommand{\hs}{\hspace*{0.3cm}}

\newcommand{\be}{\begin{equation}}
\newcommand{\ee}{\end{equation}}
\newcommand{\bea}{\begin{eqnarray}}
\newcommand{\eea}{\end{eqnarray}}
\newcommand{\ben}{\begin{enumerate}}
\newcommand{\een}{\end{enumerate}}
\newcommand{\bit}{\begin{itemize}}
\newcommand{\eit}{\end{itemize}}
\newcommand{\bde}{\begin{widetext}}
\newcommand{\ede}{\end{widetext}}
\newcommand{\nn}{\nonumber}
\newcommand{\crn}{\nonumber \\}

\newcommand{\al}{\alpha}
\newcommand{\la}{\lambda}

\newcommand{\ga}{\gamma}
\newcommand{\va}{\varphi}

\newcommand{\pa}{\partial}

\newcommand{\fr}{\frac}

\newcommand{\bc}{\begin{center}}
\newcommand{\ec}{\end{center}}

\newcommand{\ep}{\epsilon}

\newcommand{\si}{\sigma}

\newcommand{\eq}{\eqref}
\newcommand{\mathsym}[1]{{}}
\newcommand{\gev}{~\mathrm{GeV}}

\definecolor{bostonuniversityred}{rgb}{0.8, 0.0, 0.0}

\def\gsim{\raise0.3ex\hbox{$\;>$\kern-0.75em\raise-1.1ex\hbox{$\sim\;$}}}
\def\lsim{\raise0.3ex\hbox{$\;<$\kern-0.75em\raise-1.1ex\hbox{$\sim\;$}}}

\newcommand{\iop}{ Institute of Physics, Vietnam Academy Science and Technology,\\ 10 Dao Tan, Giang Vo, Hanoi 10000, Vietnam}

\newcommand{\stai}{ Subatomic Physics Research Group,
		Science and Technology Advanced Institute,\\
		Van Lang University, Ho Chi Minh City, Vietnam}
\newcommand{\steh}{  Faculty of Applied Technology, School of  Technology,  Van Lang University, Ho Chi Minh City, Vietnam}

\begin{document}
	\title{Form of axion in DFSZ models with $U(1)_{PQ}$ symmetry}
		\author{V. H. Binh$^{a,b}$}
	\email{vhbinh@iop.vast.vn }%(corresponding author)	
\author{H. N. Long$^{c,d}$}
	\email{hoangngoclong@vlu.edu.vn(corresponding author) }
	\affiliation{
	$^a$ \iop\\
	$^b$ Bogoliubov Laboratory of Theoretical Physics, Joint Institute for Nuclear Research, Dubna, 141980 Russia\\
    $^c$ \stai \\
	  $^d$ \steh \\
}

\date{\today }

%%%%%%%%%%%%
\begin{abstract}

Form of axion state in the DFSZ model is determined by exacted diagonalizing of the mass mixing matrix of $CP$-odd sector. {After applying $PQ$ mechanism to determine $PQ$ charges of particles of the model, these $PQ$ charges are used to define the general form of axion. Both of these two methods give consistent and reasonable physical state of axion. The constraint of $PQ$ charge in $DFSZ - I$ model is also pointed out. Then, anomalous couplings of axion-photon-photon is re-examined. Triple-couplings of axion with two fermions are studied. The decay of  and axion into a pair of photon is also considered via re-normalizable interactions arising from kinetic terms of scalar fields.}%\vhb{In this models, $PQ$ charge of the singlet heavy scalar and a mixing angle of doublet scalars' $PQ$ charges  are two parameters for $U(1)_{PQ}$ symmetry. Anomalous couplings of axion arising from $U(1)_{PQ}$ global symmetry spontaneous breaking is studied in parallel with couplings of axion arising from $SU(2)_L \times U(1)_Y$ local symmetry spontaneous breaking.}

\end{abstract}
%%%%%%%%%%%%


\maketitle
\noindent

%%%%%%%%%%%%%%%%%%%%%%%%
\section{Introduction}
\label{Intro}

Over forty years ago, DFSZ models were proposed \cite{DFSZ1,DFSZ2} to deal with the Strong $CP$ problem by both of imposing $U(1)_{PQ}$ symmetry and replacing Standard Model (SM) doublet scalar by two doublet scalars as well as introducing a heavy singlet scalar to SM. {$PQ$ invariance is required for all terms of Yukawa interactions as well as terms in scalar potential of the model to determine the $PQ$ charges for all particles and the constraint of $PQ$ charges, too. By applying $PQ$ mechanism, the general form of axion is pointed out and this form is consistent with the expression of axion determined from $CP$-odd mass mixing matrix diagonalizing. Through over four decades, axion has been predicted that its physical state has only one component from the imaginary part of the complex singlet scalar which causes the $U(1)_{PQ}$ spontaneous symmetry breaking \cite{DFSZ1,DFSZ2}. In this work, it is showed that there are mixtures of scalars' components that makes physical state of axion having not only the imaginary part of the singlet scalar mentioned above but also the imaginary parts of doublet scalars causing $SU(2)_L \times U(1)_Y$ spontaneous symmetry breaking. Because of this complicated structure, $DFSZ$ axion has either anomalous couplings axion-photon-photon generated from effective $QCD$ Lagrangian or triple-couplings axion-photon-photon arisen from kinetic terms of scalar fields.}

\section{Review of the model}
\label{Review}

Base on the Standard Model (SM) gauge group $\mathrm{SU(3)_C \otimes SU(2)_L \otimes U(1)_Y}$, fermion content of DFSZ models is as follows \cite{DFSZ1,DFSZ2}
\bea
Q_L & = & \left( \begin{array}{c}
u_L\\
d_L  \\
    \end{array} \right) \sim \left(3, 2, \fr 1 3\right)\,,\crn
    \Psi_L & = & \left( \begin{array}{c}
\nu_L\\
l_L  \\
    \end{array} \right) \sim \left(1, 2, -1\right)\,,\label{j171}\\
u_R & \sim & \left(3, 1, \fr 4 3\right)\,,\hs  d_R  \sim  \left(3, 1, -
\fr 2 3\right) \,,\hs  l_R \sim  \left(1, 1, - 2\right)\,.
\nn
\eea
By replacing SM-Higgs boson by two doublets and introducing one singlet scalar \cite{DFSZ1,DFSZ2}, scalar sector of DFSZ-I model consists three below scalars
\bea
H_u  & = & \left( \begin{array}{c}
         \fr{1}{\sqrt2}\left(v_u +R_{H_u^0} +i I_{H_u^0} \right)  \\
H_u^-
    \end{array} \right)  \sim \left(1, 2, - 1\right) \,, \notag\\
    H_d   &=&  \left( \begin{array}{c}
H_d^+\\
         \fr{1}{\sqrt2}\left(v_d +R_{H_d^0} +i I_{H_d^0} \right)  \\
    \end{array} \right) \sim \left(1, 2, 1\right)\,,
 \label{ScalarDoublets}\\
 \Phi & = &  \fr{1}{\sqrt2}\left(v_\phi +R_\phi +i I_\phi \right) \sim \left(1, 1, 0\right)\,. \label{ScalarSinglet}
%H_u & = & \left( \begin{array}{c}
%H_u^0\\
%H_u^- \\
%    \end{array} \right) \sim \left(1, 2, - 1\right)\,,\crn
% H_d & = & \left( \begin{array}{c}
%H_d^+\\
%H_d^0 \\
%    \end{array} \right) \sim \left(1, 2, 1\right)\,,\label{j172}\\
%\Phi & \sim & \left(1, 1, 0\right)\,.
%\nn
\eea

where the vacuum expectation values (VEVs) $ v_\phi \sim 10^{12} \text{GeV}$ \cite{vphi1,vphi2} and $v_d, v_u$ are at electroweak (EW) scale that satisfy
$v_d^2 + v_u^2 = 246^2 \text{GeV}^2$. With these VEVs, the gauge symmetry group of the model is broken through 2 stages as
\bea
SU(3)_C \times SU(2)_L \times U(1)_Y &\times& U(1)_{PQ}\notag\\
&\downarrow& \, v_\phi \sim 10^{12} \text{GeV}\notag\\
SU(3)_C \times SU(2)_L &\times& U(1)_Y\notag\\
&\downarrow& \, v_u, v_d \sim \mbox{EW scale}\notag\\
SU(3)_C &\times& U(1)_Q \label{breakingstages}
\eea
On the other hand, the scalar sector of DFSZ-II model consists not only two doublets and one singlet but also an anti-doublets \cite{DFSZ2}
%\bea
%H_u & = & \left( \begin{array}{c}
%H_u^0\\
%H_u^- \\
%    \end{array} \right) \sim \left(1, 2, - 1\right)\,,\hs
\bea
\tilde{H}_u= i \si_2 H_u^* \, =  \left( \begin{array}{c}
H_u^+\\
\fr{1}{\sqrt2}\left(v_u +R_{H_u^0} -i I_{H_u^0} \right)  \\
    \end{array} \right) \sim \left(1, 2,  1\right)\,,
%    \crn
% H_d & = & \left( \begin{array}{c}
%H_d^+\\
%H_d^0 \\
%    \end{array} \right) \sim \left(1, 2, 1\right)\,,\hs
% \Phi  \sim  \left(1, 1, 0\right)\,,\nn
\eea
with $\sigma_2$ is the second Pauli matrix.\\
With such scalar content, the Lagrangian of scalar sector is generally given by \cite{DFSZ1,DFSZ2,r1}
\bea
\mathcal{L}_{scalar} & = & \fr 1 2  \pa_\mu\Phi \pa^\mu\Phi
+(D_\mu H_d)^{\dag}D^{\mu} H_d +(D_\mu H_u)^{\dag}D^{\mu} H_u
 -V(\Phi,H_u,H_d)\,,
 \label{LagrangianScalars}
\eea
with the covariant derivative is defined as
\be
    D_\mu =  \pa_\mu - i\, g\, T_a W^a_\mu\,  - i\, g^\prime\, \fr Y 2 B_\mu \equiv \pa_\mu - i P_\mu\,   \,,
     \label{j175}
\ee
where $W_\mu^a$ and $B_\mu $ are the $SU(2)_L$ and $U(1)_Y$ gauge fields, respectively. And  $g$ and $g^\prime$  are  denoted  for coupling constants of $SU(2)_L$ and $U(1)_Y$ groups, while $T_a = \frac{\sigma_a}{2}$ with $(a=1,2,3)$, are generators of $SU(2)_L$ gauge group and $\sigma_a$ are Pauli matrices.\\
%%%%%%%%
And as usual, the electric charge operator is defined as
\be
Q = T_3 + \fr Y 2 \,,
\label{chargeoperator}
\ee
in which $T_3$ is generator of $SU(2)_L$ gauge group and $Y$ is hyper-charge of $U(1)_Y$ gauge group.
%%%%%%%%%%%%
{\subsection{Gauge sector}\label{GaugeSector}
From the kinetic term of scalar fields in Eq.~\eq{LagrangianScalars}, four electroweak gauge bosons are $W_\mu^i$, $(i=1,2,3)$ of $SU(2)_L$ group and $B_\mu$ of $U(1)_Y$ group help to determine the charge and neutral currents as
\bea
P_\mu = \frac{g}{2}\left( \begin{array}{cc}
    W_\mu^3 + \tan \theta_W \, B_\mu\, Y  & W_\mu^1 - i W_\mu^2 \\
    W_\mu^1 + i W_\mu^2 & -W_\mu^3 + \tan \theta_W \, B_\mu\, Y
\end{array}
\right)\,, \label{Pmu}
\eea
with Weinberg angle defined as $\tan \theta_W = \frac{g^\prime}{g}$ and values of $Y$ are hyper-charges of doublet scalars. For each doublet scalars, these currents are defined by replacing their own $Y$-charges into Eq.~\eq{Pmu}. % as below
%\bea
%P_\mu H_u &=& \frac{g}{2}\left( \begin{array}{cc}
%    W_\mu^3 - \tan \theta_W \, B_\mu  & W_\mu^1 - i W_\mu^2 \\
%    W_\mu^1 + i W_\mu^2 & -W_\mu^3 - \tan \theta_W \, B_\mu
%\end{array}
%\right)H_u\,, \label{PmuHu}\\
%P_\mu H_d &=& \frac{g}{2}\left( \begin{array}{cc}
%    W_\mu^3 + \tan \theta_W \, B_\mu  & W_\mu^1 - i W_\mu^2 \\
%    W_\mu^1 + i W_\mu^2 & -W_\mu^3 + \tan \theta_W \, B_\mu
%\end{array}
%\right)H_d\,, \label{PmuHd}
%\eea
As the results, the mass eigenstates of charge gauge bosons are
\bea W_\mu^\pm = \frac{1}{\sqrt{2}}\left( W_\mu^1 \mp i W_\mu^2\right)\,, \label{PhysCG}
\eea
with their corresponding masses
\bea
m_W^2 = \frac{g^2}{4}\left(v_u^2 + v_d^2 \right)\,,
\eea
while the physical state of neutral gauge boson are defined as
\bea
\left(\begin{array}{c} Z_\mu \\ A_\mu
\end{array} \right) = \left( \begin{array}{cc}
   \cos \theta_W & -\sin \theta_W  \\
   \sin \theta_W   & \cos \theta_W
\end{array} \right) \left( \begin{array}{c} W_\mu^3 \\ B_\mu
\end{array}\right)\,,\label{PhysNG}
\eea
where $A_\mu $ is a massless photon and $Z_\mu$ is a massive neutral gauge boson with mass as
\be
m_Z^2 = \frac{g^2}{4} \frac{v_d^2 +v_u^2}{\cos^2 \theta_W} = \frac{m_W^2}{\cos^2 \theta_W}\,. \label{mZmW}
\ee
These results are consistent with predictions of SM theory.
}

\subsection{Higgs potential}
\label{Hpot}

The last term of Eq.~(\ref{ScalarPotential}) is scalar potential and given by
\bea	
    V(\Phi, H_u, H_d) & = & \mu^2_\phi\Phi^* \Phi + \mu^2_{u} H_u^\dag H_u + \mu^2_{d} H_d^\dag H_d +   \la_{\phi} (\Phi^* \Phi)^2 + \la_u (H_u^\dag H_u)^2 + \la_d (H_d^\dag H_d)^2\crn
&&+ (\la_{1} H_u^\dag H_u + \la_{2} H_d^\dag H_d)(\Phi^* \Phi) +
\la_3 (H_u^\dag H_d)(H_d^\dag H_u) + \la_4 (\ep_{ij}H_u^i H_d^j)^*(\ep_{km}H_u^k H_d^m)\crn
&& +  (\la_5 (\ep_{ij}H_u^i H_d^i) (\Phi^* \Phi^*) + H.c.)
 \label{ScalarPotential}
\eea
{ Because of the hierarchy of VEVs $v_\phi \sim 10^{12} \text{GeV} \gg v_u, v_d \sim$ EW scale, from Eq.~(\ref{ScalarPotential}), it is reasonable to assume that the coupling constants of scalars at the same energy scale are much larger than the coupling constants of scalars at different energy scale as below
\bea
&&\lambda_u \approx \lambda_d \approx\lambda_3 \approx \lambda_4\,,\notag\\
&& \lambda_1 \approx \lambda_2 \approx \lambda_5\,. \label{assumelambda}
\eea}
 Then, the scalar potential stability conditions at tree level are defined as
\begin{eqnarray}
    2\mu_\phi^2 + 2 \lambda_\phi v_\phi^2 + \lambda_1 v_u^2 + \lambda_2 v_d^2  + 2 \lambda_5 v_u v_d  &=& 0 \,, \notag \\
    2 \mu_u^2 v_u+ 2\lambda_u v_u^3 + \lambda_1 v_\phi^2 v_u+ \lambda_4 v_d^2 v_u + \lambda_5 v_d  v_\phi^2 &=& 0 \,, \notag \\
    2 \mu_d^2 v_d + 2\lambda_d v_d^3 + \lambda_1 v_\phi^2 v_d +  \lambda_4 v_u^2 v_d + \lambda_5 v_u v_\phi^2 &=& 0 \,. \label{muconditions}
\end{eqnarray}
With  $\mu_\phi, \mu_u, \mu_d$ satisfy Eqs.~\eq{muconditions}, substituting these mass parameters into the scalar potential in Eq.~\eq{ScalarPotential},  the scalar mass mixing matrices are determined to dissect the physical states of scalars in detail.
{
	\subsection{Scalar sector}
	\subsubsection{Charged scalars}\label{chargedscalars}
	 The mass mixing matrix in the basis $(H_d^\pm, H_u^\pm)$, is
	\bea
	M_c^2= -\frac{\lambda_5 v_\phi^2 +(\lambda_4-\lambda_3)v_u v_d}{2}
	\left(
	\begin{array}{cc}
		\frac{v_u}{v_d} & 1 \\
	1	& \frac{v_d}{v_u}\\
	\end{array}
	\right). \label{Mcharged}
	\eea
	The matrix $M_c^2$ in Eq.(\ref{Mcharged}) gives two eigenvalues which are corresponding to a massless Goldstone boson $G_{W^\pm}$ eaten by $W^\pm$ boson and a massive charged scalar boson $H^\pm$ with mass as
    \bea
    m^2_{H^\pm} = \frac{\left(v_d^2+v_u^2\right) }{2 } \left( \lambda_3-\lambda_4-\frac{\lambda_5 v_\phi^2}{v_d v_u}\right)\,. \label{masscharge}
    \eea
     The physical states of $G_{W^\pm}$ and $H^\pm$ are given by
	\bea
    G_{W^\pm} &=& H_d^\pm  \cos \xi + H_u^\pm\sin \xi \,,\notag\\
    H^\pm &=& -H_d^\pm \sin \xi  + H_u^\pm\cos \xi \,, \label{physchargeboson}
	\eea
	with the mixing angle is defined as
	\bea
	\tan \xi = \frac{2 v_u v_d}{v_u^2 -v_d^2}\,.
	\eea
    The square of mass in Eq.~\eq{masscharge} must be positive so that a constraint for the quadric coupling of three scalar is
    \bea
    \lambda_5 \, < \left(\lambda_3 - \lambda_4 \right)\frac{v_u v_d}{v_\phi^2}\,. \label{lambda5condition}
    \eea
    From Eq.~\eq{lambda5condition}, it follows $ \lambda_5 \propto (\lambda_3 - \lambda_4) \frac{v_{EW}^2}{v_\phi^2}$ while $\lambda_3, \lambda_4$ are the coupling constants of light doublet scalars at EW scale. Because  $v_\phi \sim 10^{12} \text{GeV}$, value of $\lambda_5 \propto 10^{-12}$ is tiny that yields the mixtures between light and heavy scalars are tiny and mass of charged scalar $H^\pm $ is at TeV scale. % It is similarly in comparison with Yukawa coupling responds to proton stability in the supersymmetric 3-3-1 model \cite{longpal}.

	\subsubsection{CP-odd scalars}\label{CPoddscalars}
	The mass mixing matrix in the basis $(I_\phi, I_{H_u^0}, I_{H_d^0})$  is determined as
	\bea
	M_{odd}^2=-\lambda_5 %v_d v_u v_\phi^2
    \left(
	{\begin{array}{ccc}
    2 v_d v_u & v_d v_\phi & v_u v_\phi \\
     v_d v_\phi & \frac{v_d v_\phi^2}{ 2 v_u} & \frac{v_\phi^2}{2} \\
     v_u v_\phi & \frac{v_\phi^2}{2}  & \frac{v_u v_\phi^2}{2 v_d}
		%\frac{2}{v_\phi^2} & \frac{1}{v_u v_\phi} & \frac{1}{v_d v_\phi}\\
        %\frac{1}{v_u v_\phi} & \frac{1}{2 v_u^2} & \frac{1}{2 v_d v_u} \\
        %\frac{1}{v_d v_\phi} & \frac{1}{2 v_d v_u} & \frac{1}{2v_d^2}
	\end{array}}
	\right)\,. \label{Modd}
	\eea
	The matrix $M_{odd}^2$ in Eq.(\ref{Modd}) gives three eigenvalues. One of these corresponds to the mass of a massive pseudoscalar $A_5$ with value as
    \be
    m_{A_5}^2 =-\frac{\lambda_5}{2}\left(\frac{v_u v_\phi^2}{v_d} + \frac{v_d v_\phi^2}{v_u} + 4v_u v_d \right). \label{mA5}
    \ee
    Because $m_{A_5}^2$ must be positive and $v_d, v_u>0$, value of $\lambda_5$ must be negative. Combine with the condition in Eq.~\eq{lambda5condition}, to make $\lambda_5$ be definitely negative, it is requested that $\lambda_3 \geq \lambda_4$. \\
    Two other  eigenvalues are zero that correspond to a massless Goldstone boson $G_Z$ eaten by $Z$ boson and a massless Nambu-Goldstone boson - axion $a$, which can be a cold dark matter candidate. The physical states of three respective pseudo scalars are given by
    \bea
	\left(\begin{array}{c}
		a \\ G_Z \\  A_5
	\end{array}\right) = \left(
\begin{array}{ccc}
 \cos \beta_2 & -\sin \beta_1 \sin \beta_2 & \cos \beta_1 \sin \beta_2 \\
 0 & \cos \beta_1 & \sin \beta_1 \\
 -\sin \beta_2 &  -\sin \beta_1 \cos \beta_2 & \cos \beta_1 \cos \beta_2 \\
\end{array}
\right)
	\left(\begin{array}{c}
		I_\phi \\ I_{H_u^0} \\ I_{H_d^0}
	\end{array}\right)\,,
\label{Oddphysicalstates}
	\eea
	with the mixing angles are defined as
	\bea
	\tan \beta_1 = - \frac{v_d}{v_u}\,, \hs \tan \beta_2
    = - \frac{2 v_d}{ v_\phi \sqrt{1+ \frac{v_d^2}{v_u^2}}} %= -\frac{2v_d |\cos \beta_1|}{v_\phi }
    \,.\label{oddangles}
	\eea
Hence, the imaginary parts of scalar fields are presented via physical states of pseudo-scalars as
\bea
I_\phi &=& a\,\cos \beta_2 -  A_5 \sin \beta_2 \,, \label{Iphi}\\
I_{H_u^0} &=&  -a \, \sin \beta_1 \sin \beta_2 + G_Z \, \cos \beta_1  - A_5 \, \cos \beta_2 \sin \beta_1 \,,\label{IU}\\
I_{H_d^0} &=&  a  \cos \beta_1 \sin \beta_2+ G_Z \, \sin \beta_1 + A_5 \,  \cos \beta_1\cos \beta_2  \,. \label{ID}
\eea
In Eqs.~\eq{IU}, \eq{ID}, the expression of $I_{H_u^0}, I_{H_d^0}$ contains three physical states including axion $a$. Therefore, looking at the Yukawa interactions in Eq.~\eq{yuk1}, it is easy to see that axion $a$ has Yukawa interactions with all SM charged fermions. The coupling constants of these interactions depend on the tiny mass mixing angle $\beta_2$. With the hierarchy of VEVs $v_\phi \gg v_u, v_d$, value of the mass mixing angle $\tan \beta_2 \approx \sin \beta_2 \approx \beta_2 \sim 10^{-10}$ rad. In this scenario, all components which are related to $\sin \beta_2$ are able to be neglected. Axion contains only one imaginary component $I_\phi$ of the singlet scalar causing Strong $CP$ violation. This expression is convenient with results in Refs.\cite{DFSZ1,DFSZ2}. Structure of Goldstone boson $G_Z$ is unchanged while the expression of new pseudo scalar $A_5$ includes two imaginary components $I_{H_u^0}, I_{H_d^0}$ coming from two doublet scalars. In this case, structure of $A_5$ hints that mass of $A_5$ can be estimated at EW scale to subTeV scale.

	\subsubsection{CP-even sector}
	The mass mixing matrix in the basis $(R_\phi, R_{H_u^0}, R_{H_d^0})$  is defined as below
	\bea
	M_{even}^2 =  v_\phi\left(
	\begin{array}{ccc}
		2 \lambda_\phi  v_\phi & \lambda_5 v_d+\lambda_1 v_u & \lambda_2 v_d+\lambda_5 v_u \\
		\lambda_5 v_d+\lambda_1 v_u & 2 \frac{\lambda_u v_u^2}{v_\phi}-\frac{\lambda_5 v_d v_\phi}{2 v_u} & \frac{\lambda_4 v_d v_u}{v_\phi}+\frac{\lambda_5 v_\phi}{2} \\
		\lambda_2 v_d+\lambda_5 v_u & \frac{\lambda_4 v_d v_u}{v_\phi}+\frac{\lambda_5 v_\phi}{2} & 2 \frac{\lambda_d v_d^2}{v_\phi}-\frac{\lambda_5 v_u v_\phi}{2 v_d} \\
	\end{array}
	\right)\,.\label{Meven}
	\eea
	Because $v_\phi \gg v_u, v_d$, the terms which includes $\frac{1}{v_\phi}$ can be skipped, the mass mixing matrix $M_{even}^2$ is rewritten as
	\bea
	M^2_E= v_\phi \left(
	\begin{array}{ccc}
		2 \lambda_\phi  v_\phi & \lambda_5 v_d+\lambda_1 v_u & \lambda_2 v_d+\lambda_5 v_u \\
		\lambda_5 v_d+\lambda_1 v_u & -\frac{\lambda_5 v_d v_\phi}{2 v_u} & \frac{\lambda_5 v_\phi}{2} \\
		\lambda_2 v_d+\lambda_5 v_u & \frac{\lambda_5 v_\phi}{2} & -\frac{\lambda_5 v_u v_\phi}{2 v_d} \\
	\end{array}
	\right)\,.\label{MoE}
	\eea
	To diagonalize the mass mixing matrix $M_E^2$ in Eq.~(\ref{MoE}), the rotational matrix with two angles are defined via two stages. The first stage is dealing with $2 \times 2$ blocks below
	\bea
	M_{E_1}^2 = -\frac{\lambda_5 v_\phi^2}{2} \left(
	\begin{array}{cc}
		\frac{v_d}{v_u} & -1 \\
		-1 & \frac{v_u}{v_d} \\
	\end{array}
	\right)\,. \label{MoE1}
	\eea
	The rotational matrix which is used to diagonalize the matrix $M_{E_1}^2$ in Eq.~(\ref{MoE1}) is
	\bea
	U_{E_1}= \left( \begin{array}{cc}
		\cos \alpha_1 & \sin \alpha_1 \\
		-\sin \alpha_1 & \cos \alpha_1
	\end{array}
	\right)\,,
	\eea
	with the mixing angle is given by
	\bea
	\tan \alpha_1 = \frac{v_d}{v_u}=-\tan \beta_1,\label{alpha1}
	\eea
	and gives two eigenvalues which are zero and $m^2_H = -\frac{\lambda_5 v_\phi^2 }{2} \left(\frac{v_u}{v_d} + \frac{v_d}{v_u^2} \right)$.
    Mass split of the massive field $H$ and the massive pseudoscalar $A_5$ is of order of EW scale as below
	\bea
	 m^2_{A_5} -m^2_H  = -2\lambda_5 v_d v_u \sim \mathcal{O}_{(EW)}\,.	\eea
	Then we have the $3 \times 3$ matrix which can diagonalize a part of the mass mixing $M^2_E$
	\bea
	U_{E_1}^{3 \times 3}
	= \left(
	\begin{array}{ccc}
		1 & 0 & 0 \\
		0 &  \frac{v_u}{\sqrt{v_u^2 + v_d^2}} & \frac{v_d}{\sqrt{v_u^2 + v_d^2}} \\
		0 & - \frac{v_d}{\sqrt{v_u^2 + v_d^2}} &  \frac{v_u}{\sqrt{v_u^2 + v_d^2}} \\
	\end{array}
	\right)\,. \label{UE33}
	\eea
	%Under the effect of t
    The rotational matrix $U_{E_1}^{3 \times 3}$ in Eq.~(\ref{UE33}) effects on the mass mixing matrix $M^2_E$ and this mass mixing matrix is changed into new form as
	\bea
	M_{E^\prime}^2 	&=& v_\phi \left(
	\begin{array}{ccc}
		2 \lambda_\phi  v_\phi & \frac{\lambda_2 v_d^2+\lambda_1
			v_u^2 +2 \lambda_5 v_d v_u}{\sqrt{v_u^2 + v_d^2}} & \frac{\lambda_5 \left( v_u^2 -v_d^2 \right)+ (\lambda_2-\lambda_1)v_d v_u}{\sqrt{v_u^2 + v_d^2}} \\
		 \frac{\lambda_2 v_d^2+\lambda_1
			v_u^2 +2 \lambda_5 v_d v_u}{\sqrt{v_u^2 + v_d^2}} & 0 & 0 \\
		\frac{\lambda_5 \left( v_u^2 -v_d^2 \right)+ (\lambda_2-\lambda_1)v_d v_u}{\sqrt{v_u^2 + v_d^2}} & 0 & -\frac{\lambda_5 v_\phi }{2}\left(\frac{v_u}{v_d}+\frac{v_d}{v_u}\right) \\
	\end{array}
	\right)\,.\label{MoE2}
	\eea
   To make it simpler, the third element of the first row (as well as the third element of the first column) of the matrix $M^2_{E^\prime}$ is assumed to be suppressed. As the result, there is a constraint as
   \bea
    \lambda_5 = \frac{v_u v_d}{v_u^2 -v_d^2}\left( \lambda_1 - \lambda_2 \right)\,. \label{lambda521}
   \eea
   This contraint in Eq.~(\ref{lambda521}) is consistent with the assumption in Eq.~(\ref{assumelambda}) that is mentioned in subsection \ref{Hpot}.
	 Eq.~\eq{lambda521} also shows that $(\lambda_2 - \lambda_1) \sim \lambda_5 \propto 10^{-12}$ (see  \ref{chargedscalars}) and helps the matrix $	M_{E^\prime}^2$ turn out to be $M_{E^{\prime\prime}}^2$ in form of
	\bea
   M_{E^{\prime\prime}}^2 &=&  v_\phi\left(
	\begin{array}{ccc}
		2 \lambda_\phi  v_\phi & \frac{v_{EW}\left(\lambda_2 v_d^2 -\lambda_1 v_u^2\right)}{v_d^2 -v_u^2} &0 \\
        \frac{v_{EW}\left(\lambda_2 v_d^2 -\lambda_1 v_u^2\right)}{v_d^2 -v_u^2} &0 &0 \\
        0 & 0 & \frac{v_\phi v_{EW}^2 \left(\lambda_2 -\lambda_1 \right)}{2\left(v_u^2 -v_d^2 \right)}
	\end{array}
	\right)\,,
    \label{MoE2s}
	\eea
   % with $v_{EW}^2 = v_d^2 + v_u^2 = 246^2 \gev^2$ is at EW scale.
Now, the second stage is dealing with the $2 \times 2$ first block of $M_{E^{\prime\prime}}^2$. This matrix is diagonalized by the below rotational matrix
\bea
U_{E_2}= \left( \begin{array}{cc}
	\cos \alpha_2 & \sin \alpha_2 \\
	-\sin \alpha_2 & \cos \alpha_2
\end{array}
\right)\,, \label{UE2}
\eea
with the mixing angle defined as
\bea
\tan 2 \alpha_2 = \frac{v_{EW}\left(\lambda_2 v_d^2 - \lambda_1 v_u^2 \right)}{ \lambda_\phi v_\phi \left( v_d^2 -v_u^2\right)}\,.\label{alpha2}
\eea
The $2 \times 2$ first block of matrix $M^2_{E^{\prime\prime}}$ gives two eigenvalues as below
\bea
m^2_{H^\prime} &=& 2 \lambda_\phi v_\phi^2 + \frac{v_{EW}^2}{2 \lambda_\phi}\left(\frac{\lambda_1 v_u^2 -\lambda_2 v_d^2}{v_d^2 -v_u^2}\right)^2\,,\label{mH1}\\
m^2_{h} &=& -\frac{v_{EW}^2}{2 \lambda_\phi}\left(\frac{\lambda_1 v_u^2 -\lambda_2 v_d^2}{v_d^2 -v_u^2}\right)^2 \,. \label{mSMLHiggs}
\eea
From Eq.~(\ref{mSMLHiggs}), it is shown that the scalar $h$ gets mass at EW scale when $\lambda_\phi<0$ and its absolute value is $|\lambda_\phi| \sim \lambda_1^2, \lambda_2^2$. Reminding to Eq.~\eq{lambda521}, it is shown that $\lambda_1, \lambda_2 \sim \lambda_5 \propto 10^{-12}$. The absolute value of $\lambda_\phi$ should be around $10^{-6}$ which is consistent with the assumption in Eq.~(\ref{assumelambda}). Then, the massive scalar $h$ is the SM-like Higgs of the model under consideration. Mass of $H$ field is $m_H^2 = \frac{(\lambda_1 -\lambda_2)v_{EW}^2 v_\phi^2}{2(v_d^2 -v_u^2)}$ at TeV scale, while  $H^\prime $ is very heavy  with mass at $v_\phi$ scale as given in Eq.~\eq{mH1}.
The mass mixing matrix $M^2_{E}$ in Eq.~(\ref{MoE}) is diagonalized by the $3 \times 3$ rotational matrix
\bea
U_{even} =
\left(
\begin{array}{ccc}
 \cos \alpha_2 & \sin \alpha_2 & 0 \\
 -\cos \alpha_1 \sin \alpha_2 & \cos \alpha_1 \cos
   \alpha_2 & \sin \alpha_1 \\
 \sin \alpha_1 \sin \alpha_2 & -\sin \alpha_1 \cos
   \alpha_2 & \cos \alpha_1 \\
\end{array}
\right)
	%\cos \alpha_2 & \cos \alpha_1 \sin \alpha_2 & \sin \alpha_1 \sin \alpha_2 \\
	%-\sin \alpha_2 & \cos \alpha_1 \cos \alpha_2 & \sin \alpha_1 \cos \alpha_2 \\
	%0 & -\sin \alpha_1 & \cos \alpha_1 \\
\,.
\eea
Approximately, the mass mixing matrix $M^2_{even}$ is diagonalized and gives three non-zero eigenvalues $  m_{H^\prime}^2, m_{h}^2, m_H^2$. The physical fields are presented via $R_\phi, R_{H_u^0}, R_{H_d^0}$ as
\bea
\left(\begin{array}{c}
	H^\prime \\ h \\ H
\end{array}\right) = \left(
\begin{array}{ccc}
\cos \alpha_2 & \sin \alpha_2 & 0 \\
 -\cos \alpha_1 \sin \alpha_2 & \cos \alpha_1 \cos
   \alpha_2 & \sin \alpha_1 \\
 \sin \alpha_1 \sin \alpha_2 & -\sin \alpha_1 \cos
   \alpha_2 & \cos \alpha_1 \\
%\cos \alpha_2 & \cos \alpha_1 \sin \alpha_2 & \sin \alpha_1 \sin \alpha_2 \\
%-\sin \alpha_2 & \cos \alpha_1 \cos \alpha_2 & \sin \alpha_1 \cos \alpha_2 \\
%0 & -\sin \alpha_1 & \cos \alpha_1 \\
\end{array}
\right)\left(\begin{array}{c}
R_\phi \\R_{H_u^0} \\R_{H_d^0}
\end{array}\right)\,.
\label{may221}
\eea
Because of the hierarchy $v_\phi \gg v_{EW}$, the mixing angle $\alpha_2$ given in Eq.~\eq{alpha2} is tiny so that $\sin \alpha_2$ can be neglected. In this scenario, the physical states of $CP$-even scalar fields are represented as
\bea
H^\prime &=& \cos \alpha_2 \, R_\phi\,,\notag \\
h &=& \cos \alpha_1 \cos \alpha_2 \, R_{H_u^0} + \sin \alpha_1 \, R_{H_d^0} \,, \notag\\
H &=& -\sin \alpha_1 \cos \alpha_2\, R_{H_u^0} + \cos \alpha_1 \, R_{H_d^0}\,.\label{RealPhysScalar}
\eea
The expressions in Eq.~(\ref{RealPhysScalar}) truly show that $h$ and $H$ with two imaginary components coming from doublet scalars with VEVs at EW scale and there are two light scalar bosons in DFSZ models, one of them is SM-like Higgs boson ($h$) and the another is a new light scalar $H$. Besides, $H^\prime$ with only one imaginary from singlet scalar with VEV at $10^{12}$ GeV, is a super heavy scalar boson which can be used to consider the inflation of Early Universe.
%And the real parts of scalar fields are presented via physical states of scalars as below
%\bea
%R_\phi &=& \cos \alpha_2 \, H^\prime\,,\notag\\
%R_{H_u^0} &=& \cos \alpha_1 \cos \alpha_2 \, h  - \sin \alpha_1  \, H\,,\notag\\
%R_{H_d^0} &=& \sin \alpha_1 \cos \alpha_2\, h  + \cos \alpha_1 \, H\,.\label{RealGaugeScalar}
%\eea
%Replacing these expressions in Eq.~(\ref{RealGaugeScalar}) into Yukawa interactions in Eq.~(\ref{YukDFSZ1}), we can study on decays of SM-like Higgs into a pair of fermions (e.g. a pair of $b$ quarks or a pair of charged leptons).

\section{Yukawa interactions}
\subsection{Yukawa interactions in DFSZ-I model}\label{YukI}
With the spectrum of particles introduced in Sec.~\ref{Review}, the Yukawa interactions in DFSZ-I model are determined as
\bea
	\mathcal{L}_{Yukawa}^{DFSZ-I} & = &	y_u^{H_u}\, \overline{Q}_L\, H_u\, u_R + y_d^{H_d}\, \overline{Q}_L\, H_d\, d_R + y_l^{H_d}\, \overline{\Psi}_{L}\, H_d\, l_R + \text{H.c.}\label{yuk1}\\
& = &	y_u^{H_u}\, (\overline{Q}_L\, H_u\, u_R + \overline{u}_R\, H_u^\dag\, Q_L) + y_d^{H_d}\,( \overline{Q}_L\, H_d\, d_R + \overline{d}_R\, H_d^\dag\, Q_L) +  y_l^{H_d}\, (\overline{\Psi}_{L}\, H_d\, l_R + \overline{l}_{R}\, H_d^\dag\, \Psi_L )\label{yuk2}\\
& = &	y_u^{H_u} ( \overline{u}_L H_u^{0} u_R +
\overline{d}_L H_u^- u_R + \overline{u}_R H_u^{0*} u_L + \overline{u}_R H_u^{+} d_L ) \crn  &&+ y_d^{H_d} ( \overline{u}_L  H_d^{+} d_R +
\overline{d}_L H_d^0 d_R + \overline{d}_R  H_d^{-} u_L + \overline{d}_R  H_d^{0*} d_L)\crn
&& +  y_l^{H_d}(\overline{\nu}_L H_d^{+} l_R
+ \overline{l}_L H_d^{0} l_R + \overline{l}_R H_d^{-} \nu_L + \overline{l}_R H_d^{0*} l_L )\,,
\label{YukDFSZ1}
\eea
where $(\overline{a}_L\, H\, b_R)^\dag = \overline{b}_R\, H^\dag \, a_L$ was used.

\subsection{Yukawa interactions in DFSZ-II model}\label{YukII}
In DFSZ-II model, the particle spectrum was reviewed in Sec.~\ref{Review}, %includes all particles of DFSZ-I model and an anti-doublet scalar \cite{DFSZ2}
%\bea
%\tilde{H}_u= i \si_2 H_u^* \, =  \left( \begin{array}{c}
%H_u^+\\
%H_u^{0 *} \\
%    \end{array} \right) \sim \left(1, 2,  1\right)\,, \label{antidoublet}
%\eea
%with the neutral component is  expanded around its VEV as
%\bea
%H_u^{0 *} = \fr{1}{\sqrt2}\left(v_u +R_{H_u^0} - i I_{H_u^0} \right)\,. \label{antiHu}
%\eea
Lagrangian of Yukawa interactions in DFSZ-II model is given as below
\bea
\mathcal{L}_{Yukawa}^{DFSZ-II} & = &	y_u^{H_u}\, \overline{Q}_L\, H_u\, u_R + y_d^{H_d}\, \overline{Q}_L\, H_d\, d_R %+y_l^{H_d}\, \overline{\Psi}_{L}\, H_d\, l_R
+ y_l^{H_u}\, \overline{\Psi}_{L}\, \title{H}_u\, l_R + \text{H.c.}\crn
& = &	y_u^{H_u}\, (\overline{Q}_L\, H_u\, u_R + \overline{u}_R\, H_u^\dag\, Q_L) + y_d^{H_d}\,( \overline{Q}_L\, H_d\, d_R + \overline{d}_R\, H_d^\dag\, Q_L) \crn
&& +   y_l^{H_u}\, (\overline{\Psi}_{L}\, \tilde{H}_u\, l_R + \overline{l}_{R}\, \tilde{H}_u^\dag\, \Psi_L )%+ y_l^{H_d}\,( \overline{\Psi}_L\, H_d\, l_R + \overline{l}_R\, H_d^\dag\, \Psi_L)\,.
\label{YukDFSZ2a}\\
&=& y_u^{H_u} ( \overline{u}_L H_u^{0} u_R + \overline{d}_L H_u^- u_R + \overline{u}_R H_u^{0*} u_L + \overline{u}_R H_u^{+} d_L ) \crn  &&+ y_d^{H_d} ( \overline{u}_L  H_d^{+} d_R + \overline{d}_L H_d^0 d_R + \overline{d}_R  H_d^{-} u_L + \overline{d}_R  H_d^{0*} d_L)\crn
&& +  y_l^{H_u}(\overline{\nu}_L H_u^{+} l_R + \overline{l}_L H_u^{0*} l_R + \overline{l}_R H_u^{-} \nu_L + \overline{l}_R H_u^{0} l_L )\,.
%&& +  y_l^{H_d}(\overline{\nu}_L H_d^{+} l_R + \overline{l}_L H_d^{0} l_R + \overline{l}_R H_d^{-} \nu_L + \overline{l}_R H_d^{0*} l_L )
\label{YukDFSZ2}
\eea
There are differences in terms of Yukawa interactions of leptons. Hence, masses of charged leptons are generated by neutral electrical component of $H_u$ instead of the one of $H_d$ as in DFSZ-I model.
}
\section{$PQ$ charges}\label{PQchargeSec}
%Besides defining the symmetries of the model, we assume that the classical Lagrangian automatically possesses the global Peccei-Quinn symmetry.
%In Ref.\cite{julio2},
 The rules to determine $PQ$ charges of multiplets and fields of any Beyond Standard Models (BSMs) which include $U(1)_{PQ}$, were introduced in Ref.~\cite{r2}.
Remind that the $PQ$ charge of a multiplet is an average value of all components' $PQ$ charges in the multiplet.  Moreover,  behaviors of fermion and scalar fields under $U(1)_{PQ}$ transformation are
\bea
f & \rightarrow & f^\prime =  e^{ i  \al \ga_5 PQ_{(f)} } f\,, \hs {\bar f} \rightarrow {\bar f}^\prime =  {\bar f} e^{ i  \al \ga_5 PQ_{(f)} } \,, \hs
\va \rightarrow \va^\prime = e^{ - i  \al  PQ_{(\va)} } \va \,, \label{au311}\\
f_R &\rightarrow & f^\prime_R  =  e^{  i \al \left[PQ_{(f)} \right] } f_R\,,\hs
{\bar f}_R \rightarrow {\bar f}^\prime_R  = {\bar f}_R  e^{ - i \al  \left[PQ_{(f)} \right] }\,,\crn
 f_L &\rightarrow& f^\prime_L  =  e^{ - i \al  \left[PQ_{(f)} \right]} f_L\,,\hs
{\bar f}_L \rightarrow {\bar f}^\prime_L  = {\bar f}_L  e^{  i \al \left[PQ_{(f)} \right]}\,, \label{hay1n}
\eea
with $PQ_{(x)}$, $(x = f, \va)$, are the $PQ$ charges of fields and $f_L, f_R$ are left-handed and right-handed fermions, respectively.  For the convenience, the minus sign in the exponent of scalar field $\va$ is used.

%For chiral fermions, %which are often used in the literature,there is no
%$\ga_5$ disappears in the exponent of transformations as below
%\bea
% f_R &\rightarrow & f^\prime_R  =  e^{  i \al \left[PQ_{(f)} \right] } f_R\,,\hs
%{\bar f}_R \rightarrow {\bar f}^\prime_R  = {\bar f}_R  e^{ - i \al  \left[PQ_{(f)} \right] }\,,\crn
% f_L &\rightarrow& f^\prime_L  =  e^{ - i \al  \left[PQ_{(f)} \right]} f_L\,,\hs
%{\bar f}_L \rightarrow {\bar f}^\prime_L  = {\bar f}_L  e^{  i \al \left[PQ_{(f)} \right]}\,, \label{hay1n}
%\eea
%with $f_L, f_R$ are left-handed and right-handed fermions, respectively.
Note that the combination of Eqs.\eq{au311} and \eq{hay1n} helps to describe the transformation of fermion for short as %leads to the following relation
\be
PQ_{(f)} = PQ_{(f_R)} = - PQ_{(f_L)}=PQ_{(\bar{f_L})}\,.
\label{a161}
\ee
%Within the above definition, it is obviously that the $PQ$ charge of a multiplet must follow the special rules given in  Eq.~\eq{au311} or in  Eq.~\eq{hay1n} for chiral fermions.
%Therefore, t
There are {\it two}  properties of the $PQ $ charge of a multiplet \cite{r2} as
\bea PQ_{\overline{f}_L} = - PQ_{f_L}\,,\hs  PQ_{\phi^*} = - PQ_{\phi}\,.%,\hs PQ_{(S_L)^c} = - PQ_{S_L}\,.
\eea
%\ben
%\item The PQ charge of a left-handed multiplet is opposite in sign to the one of the corresponding right-handed multiplet;
%\item The PQ charge of a multiplet is opposite in sign to the one of the %corresponding anti-multiplet.
%\bea PQ_{\overline{f}_L} = - PQ_{f_L}\,,\hs  PQ_{\phi^*} = - PQ_{\phi}\,.%,\hs PQ_{(S_L)^c} = - PQ_{S_L}\,.
%\eea
%\een
To determine $PQ$ charges of fields, all terms of the Yukawa interactions and scalar potential are assumed to automatically possesses the global Peccei-Quinn symmetry.  Similarly with Ref.~\cite{r2}, the relation between $PQ$ charge of the scalar and $PQ$ of two fermions is %  must be invariant under $U(1)_{PQ}$ transformation. This condition help us to determine $PQ$ charges of all particles in the considered model.
%Hence, we start with the general Yukawa coupling of two fermions and a scalar
%\be \mathcal{L}_{Y}^f = h  {\bar f}_L \phi \psi_R + H.c.
%\label{a162}
%\ee
%Under the $PQ $ transformation, it becomes
%\be {\mathcal{L}^f_{Y}}^\prime  = h{\bar f}'_L \phi^\prime  \psi^\prime_R = h  \, e^{  i \al  PQ_{ (\bar{f}_L)} }{\bar f}_L  \,e^{ - i \al  PQ_{(\phi)}} \phi\,  e^{  i \al  PQ_{(\psi_R)} }  \psi_R + H.c.
%\label{a163}
%\ee
%The above Lagrangian must be invariant under the $PQ$ transformations, that means ${\mathcal{L}^f_{Y}}^\prime = \mathcal{L}_{Y}^f $. As the result, the $PQ$ charge of the scalar is equal to the sum of the fermion $PQ$ charges
\be PQ_{ (\phi)} =   PQ_{(\bar{f}_L)} + PQ_{(\psi_R)}\,.
\label{PQscalar_fermions}
\ee

%%%%%%%%%%%%%%
Apply the condition in Eq.~\eq{PQscalar_fermions} to all terms of Yukawa interaction in Eq.~\eq{yuk2}, the relations for $PQ$ charges of mutiplets are
\bea
PQ_{ Q_L} & = &  PQ_{H_u} + PQ_{ u_R}\,,\label{PQdoublets1}\\
PQ_{ Q_L} & = &  PQ_{H_d} + PQ_{ d_R}\,,\label{PQdoublets2}\\
PQ_{ \Psi_L} & = &  PQ_{H_d} + PQ_{ l_R}\,.\label{PQdoublets3}
\eea
From Eq.~\eq{PQdoublets1} and Eq.~\eq{PQdoublets2}, one has
\be
 PQ_{H_u} -  PQ_{H_d} = PQ_{ d_R} - PQ_{ u_R}\,.
 \label{PQdoubletscalars}
 \ee
 Combining three equations Eq.~\eq{PQdoublets1},~\eq{PQdoublets2},~\eq{PQdoublets3} helps to get
 \bea
PQ_{Q_L} -PQ_{\Psi_L} = PQ_{d_R} - PQ_{l_R} = PQ_{H_u} - PQ_{H_d} + PQ_{u_R} -PQ_{l_R}\,. \label{PQdoublets4}
 \eea
 Moreover, the scalar fields transform under $PQ $ transformation as below
\bea
H_u & \rightarrow & H_u^\prime = e^{ - i  \al  PQ_{H_u} } H_u \,,\crn
H_d & \rightarrow & H_d^\prime = e^{ - i  \al  PQ_{H_d} } H_d \,
\label{ScalarPQtransforms}\\
\Phi & \rightarrow & \Phi^\prime = e^{ - i  \al  PQ_{\Phi} } \Phi\,.
\label{PQscalarRules}
\eea
Requiring the $U(1)_{PQ}$ preservation for the last term of scalar potential in Eq.~\eq{ScalarPotential} gives
\be
2 PQ_{\Phi} = PQ_{H_u} + PQ_{H_d}\,.
\label{PQscalar}
\ee
 From Eq.~\eq{YukDFSZ1}, $PQ$ charges of scalar components are related to $PQ$ charges of fermions as below
\bea
&& PQ_{H_u^0} + PQ_{ u_R} +PQ_{ \bar{u}_L} = 0\,, \label{PQf1}\\
&& PQ_{H_u^-} +  PQ_{ u_R} + PQ_{\bar{ d}_L} = 0 \,, \label{PQf2}\\
&& PQ_{H_u^+} +  PQ_{ d_L}  +  PQ_{\bar{ u}_R} = 0 \,, \label{PQf3}\\
&& PQ_{H_d^0} +  PQ_{ d_R} +  PQ_{ \bar{d}_L} = 0 \,, \label{PQf4}\\
&& PQ_{H_d^+}+  PQ_{ d_R} +  PQ_{\bar{ u}_L}= 0 \,, \label{PQf5}\\
&&  PQ_{H_d^-} +  PQ_{ u_L} +  PQ_{\bar{ d}_R} = 0 \,, \label{PQf6}\\
&& PQ_{H_d^+} + PQ_{ l_R} +  PQ_{ \bar{\nu}_L} = 0 \,, \label{PQf7}\\
&& PQ_{H_d^-}+  PQ_{\bar{ l}_R} +  PQ_{ \nu_L} = 0 \,,  \label{PQf8} \\
&&  PQ_{H_d^0} +  PQ_{ \bar{l}_L}  +PQ_{l_R} = 0 \,.\label{PQf9}
\eea
%Solve the equations of Eq.~(\ref{PQf1},\ref{PQf2},\ref{PQf3},\ref{PQf4},\ref{PQf5},\ref{PQf6},\ref{PQf8},\ref{PQf9}) to get the below relations.
Eqs.~\eq{PQf2}, \eq{PQf3}, \eq{PQf5}, \eq{PQf6}, \eq{PQf7}, \eq{PQf8} lead to
\bea
 PQ_{H_u^-} = - PQ_{H_u^+} =   PQ_{H_d^+} = - PQ_{H_d^-} =  PQ_{d_L} - PQ_{ u_R}  = PQ_{\nu_L} - PQ_{l_R} \,, \label{PQf10}
\eea
while Eqs.~\eq{PQf1}, \eq{PQf4}, \eq{PQf9} give
\bea
PQ_{u_R} &=& - \frac{PQ_{H_u^0}}{2}\,, \label{PQf11} \\
PQ_{d_R} &=& PQ_{l_R} = - \frac{PQ_{H_d^0}}{2}\,. \label{PQf12}
\eea
Supplanting Eq.~\eq{PQf12} into Eq.~\eq{PQdoublets4} results in
\bea
&& PQ_{H_u} - PQ_{H_d} + PQ_{u_R} -PQ_{l_R} = 0 \,, \label{PQf13}\\
&& PQ_{Q_L}  = PQ_{\Psi_L} \,. \label{PQf14}
\eea
Substituting Eqs.~\eq{PQf11}. \eq{PQf12} into Eq.~\eq{PQf10} to get
\bea
PQ_{H_u^-} = - PQ_{H_u^+} =   PQ_{H_d^+} = - PQ_{H_d^-} =  \frac{PQ_{H_u^0}+PQ_{H_d^0}}{2}\,. \label{PQs1}
\eea
Substituting Eqs.~\eq{PQf11}. \eq{PQf12} into Eq.~\eq{PQdoubletscalars} to have
\bea
PQ_{H_u} - PQ_{H_d} = \frac{PQ_{H_u^0}-PQ_{H_d^0}}{2}\,. \label{PQs2}
\eea
Solve Eqs.~\eq{PQscalar}, \eq{PQs2}, one has
\bea
PQ_{H_u} &=&  PQ_\phi + \frac{PQ_{H_u^0}-PQ_{H_d^0}}{4}\,, \label{PQs3}\\
PQ_{H_d} &=&  PQ_\phi - \frac{PQ_{H_u^0}-PQ_{H_d^0}}{4}\,. \label{PQs4}
\eea
From another perspective, $PQ$ charges of doublet scalars are their average of $PQ$ charges of each components in the doublets. Hence, it follows
\bea
PQ_{H_u} &=& \frac{PQ_{H_u^0}+PQ_{H_u^-}}{2}\,, \label{PQs5}\\
PQ_{H_d} &=& \frac{PQ_{H_d^0}+PQ_{H_d^+}}{2}\,. \label{PQs6}
\eea
Substituting Eq.~\eq{PQs1} into Eqs.~\eq{PQs5}, \eq{PQs6}, there are below relations
\bea
PQ_{H_u} &=&  %\frac{PQ_{H_u^0}+\frac{PQ_{H_u^0}+PQ_{H_d^0}}{2}}{2} =
\frac{3PQ_{H_u^0}+PQ_{H_d^0} }{4}\,, \label{PQs7} \\
PQ_{H_d} &=&  %\frac{PQ_{H_d^0}+\frac{PQ_{H_u^0}+PQ_{H_d^0}}{2}}{2} =
\frac{3PQ_{H_d^0}+PQ_{H_u^0} }{4}\,, \label{PQs8}
\eea
Combine Eq.~\eq{PQs3} and Eq.~\eq{PQs7}, the $PQ$ charge correlation between singlet and two neutral electrical charged components of doublet scalar is
\bea
%&& PQ_\phi + \frac{PQ_{H_u^0}-PQ_{H_d^0}}{4} =\frac{3PQ_{H_u^0}+PQ_{H_d^0} }{4}\,, \notag\\
%\Longleftrightarrow &&
PQ_{H_u^0}+PQ_{H_d^0} =   2PQ_\phi \,. \label{PQs9}
\eea
Combine Eq.~\eq{PQs4} and Eq.~\eq{PQs8}, one also gets the result as in Eq.~\eq{PQs9}. Substituting Eq.~\eq{PQs9} into Eq.~\eq{PQs1}, the relation between $PQ$ charge of singlet scalar and two electric charged components of doublet scalars are
\bea
PQ_{H_u^-} = - PQ_{H_u^+} =   PQ_{H_d^+} = - PQ_{H_d^-} =  PQ_\phi \,.\label{PQs10}
\eea
Replacing Eq.~\eq{PQs10} into Eqs.~\eq{PQs5}, \eq{PQs6} leads to the $PQ$ charge correlation between singlet and doublet scalars is
\bea
PQ_{H_u} + PQ_{H_d} =  2 PQ_\phi\,. \label{PQs11}
\eea
Once again, replace Eq.~\eq{PQs9} into Eqs.~\eq{PQs7}, \eq{PQs8} to get the below definitions
\bea
PQ_{H_u} &=&   %\frac{3PQ_{H_u^0}+PQ_{H_d^0} }{4} = \frac{PQ_{H_u^0}}{2} + \frac{PQ_{H_u^0}+PQ_{H_d^0}}{4} =
\frac{PQ_{H_u^0}}{2} + \frac{PQ_\phi}{2}\,, \label{PQs12} \\
PQ_{H_d} &=&  %\frac{3PQ_{H_d^0}+PQ_{H_u^0} }{4} =  \frac{PQ_{H_d^0}}{2} + \frac{PQ_{H_u^0}+PQ_{H_d^0}}{4} =
\frac{PQ_{H_d^0}}{2} + \frac{PQ_\phi}{2}\,, \label{PQs13}
\eea
In the subsection \ref{CPoddscalars}, there is a mixture of $I_\phi, I_{H_u^0}, I_{H_d^0}$ that leads to three orthogonal physical states $a, G_Z, A_5$. It is reasonable to assume that the mixture between $PQ$ charges of two neutral electrical components of doublet scalars via a mixing angle $\zeta$ satisfying Eq.~\eq{PQs9}. Therefore, $PQ$ charges of $H_u^0$ and $H_d^0$ are represented via $\zeta$ and $PQ_\phi$  as
\bea
PQ_{H_u^0} &=&  2 \sin^2 \zeta\, PQ_\phi, \label{PQs14}\\
PQ_{H_d^0} &=&  2 \cos^2 \zeta\, PQ_\phi. \label{PQs15}
\eea
As the result, $PQ$ charges of particles are presented via two parameters. The first one is the mixing angle between $PQ$ charges of neutral electrical components of doublet scalars. The second one is $PQ_\phi$ of a singlet scalar which causes the $PQ$ symmetry breaking. $PQ$ charges of particles in the model are shown in Table \ref{tab1}

%%%%%%%%%%%%%%%%
%\small{
		\begin{table}[th]
			\resizebox{17cm}{!}{
				\begin{tabular}{|c|c|c|c|c|c|c|c|c|c|c|c|c|c|}
					\hline
			&$Q_L$		&  $\, u_R \, $ & $\, d_R\, $ & $\psi_L$ & $l_R$ & $\, \nu_L\, $ &$ H_u$ &  $H_u^0$ & $H_u^-$ & $H_d$ & $H_d^0 $ & $H_d^+$ & $\Phi$  \\ \hline
$U(1)_{PQ}$	& $\frac{PQ_\phi}{2}$ &  $-s^2_\zeta PQ_\phi   $ & $ -c^2_\zeta PQ_\phi $ & $ \frac{PQ_\phi}{2} $ & $ -c^2_\zeta PQ_\phi $ & $ s^2_\zeta PQ_\phi  $ &$\left( s^2_\zeta+ \frac{1}{2}\right) PQ_\phi  $ &  $ 2 s^2_\zeta PQ_\phi$ & $  PQ_\phi$ & $\left( c^2_\zeta+ \frac{1}{2}\right) PQ_\phi  $ & $ 2 c^2_\zeta PQ_\phi $ &  $ PQ_\phi $ & $ PQ_\phi$ 	 \\ \hline
		\end{tabular}}
			\caption{$U(1)_{PQ}$ charge assignments of the particle content of the model with $\sin^2_\zeta = s^2_\zeta, \cos^2_\zeta = c^2_\zeta $.}% with these notations  $PQ_F = PQ_{F_R} = - PQ_{F_L}$.}			}
			\label{tab1}
		\end{table}

\section{Constraint for $PQ$ charges}
%\subsection{Form  of axion state}
%\label{axionstate}
Let us pay attention to the last term in  Eq.~\eq{ScalarPotential}
{\bea
\mathcal{L}_{fours} &=&  \lambda_5 (\epsilon_{ij}H_u^i H_d^j) (\Phi^* \Phi^*) + \text{H.c} \notag\\
&=& \lambda_5(H_u^0H_d^0- H_u^-H_d^+)(\Phi^* \Phi^*) + \text{H.c}\,.
 \label{4scalars}
 \eea
 In Eq.~\eq{4scalars}, only the term that contains four neutral electrical scalar field ($ \propto H_d^0 H_u^0 (\Phi^*\Phi^*)$) generates axion field. Hence, in order to study on form of axion, we exactly focus on this term. To make it be simpler, the scalar field is rewritten in polar form \cite{r2} so that the value of an usual complex scalar $\Phi$ at its VEV ($v_\phi$) is defined as follow
\be  \Phi_{v_\phi} \equiv \Phi|_{R_\phi = v_\phi} =\frac{v_\phi}{\sqrt 2} e^{i \frac{I_\phi }{v_\phi}}\,.
\label{polarform}
\ee
1) Under $\mathrm{U(1)_Y}$ transformation, the neutral scalar fields transform as
\bea H_u^0 & \rightarrow & {H_u^0}^\prime = \frac{v_{u}}{\sqrt 2} e^{i Y_{H_u^0} \frac{I_{H_u^0} }{v_{u}}}\,,\label{U1Ytransform1} \\
H_d^0 & \rightarrow & {H_d^0}^\prime = \frac{v_{d}}{\sqrt 2} e^{i Y_{H_d^0} \frac{I_{H_d^0} }{v_{d}}}\,,
\label{U1Ytransform2}
\eea
and $\phi$ is unchanged because it does not carry $Y$ charge. Reminding that $Y_{H_u^0} = - Y_{H_d^0} = -1$ and demanding the interaction in Eq.~\eq{4scalars} is preserved under $U(1)_Y$ transformation, it follows
	\begin{equation}
	 -  \frac{I_{H_u^0} }{v_{u}}	+  \frac{I_{H_d^0} }{v_{d}}  =0 \,. \label{Ychargeconditions}
	\end{equation}
Therefore,  with $\tan \beta_1$ is given in Eq.~(\ref{oddangles}), one has
\be
 I_{H_d^0}  = \frac{v_d}{v_u} I_{H_u^0} = -\tan \beta_1\,  I_{H_u^0}\,.
 \label{IHdHu}
 \ee
 Eq.~\eq{IHdHu} shows that there is a mixture among imaginary parts of doublet scalars as the assumption when discussing in Sec.\ref{PQchargeSec} . Because of this mixture, axion with three components as in Eq.~\eq{Oddphysicalstates} can be represented by two imaginary components i.e $(I_\phi, I_{H_u^0})$ or $(I_\phi, I_{H_d^0})$.\\
2) From the other end of the discussion, the scalar fields transform under $U(1)_Y \times U(1)_{PQ}$ symmetry group as belows
	\bea
    {H_u^0}^\prime & \rightarrow & {H_u^0}^{\prime\prime} = \frac{v_{u}}{\sqrt 2} e^{i Y_{H_u^0} \frac{I_{H_u^0}} {v_{u}} PQ_{H_u^0}}\,,\label{U1YU1PQu} \\
    {H_d^0}^\prime & \rightarrow & {H_d^0}^{\prime\prime} = \frac{v_{d}}{\sqrt 2} e^{i Y_{H_d^0} \frac{I_{H_d^0} }{v_{d}}PQ_{H_d^0}}\,,\label{U1YU1PQd} \\
    \Phi & \rightarrow & \Phi^{\prime\prime} =  \frac{v_{\phi}}{\sqrt 2} e^{i \frac{I_\Phi}{v_\phi} PQ_\phi}\,. \label{U1YU1Pphi}
\eea
Requiring the interaction in Eq.~\eq{4scalars} is preserved under $U(1)_Y \times U(1)_{PQ}$ transformation helps to get the following constraint
    \be
   	Y_{H_u^0} \frac{I_{H_u^0} }{v_{u}}  PQ_{H_u^0} 	+  Y_{H_d^0} \frac{I_{H_d^0} }{v_{d}}PQ_{H_d^0} -
2 \frac{I_{\Phi} }{v_{\phi}}PQ_{\Phi}  =0 \,. \label{U1YU1PQ4}
	\ee
Once again, replacing $Y$ charges of scalars into Eq.~\eq{U1YU1PQ4} yields
    \be
   	\frac{I_{H_u^0} }{v_{u}} PQ_{H_u^0}	- \frac{I_{H_d^0} }{v_{d}}PQ_{H_d^0} +2 \frac{I_{\Phi} }{v_{\phi}}PQ_{\Phi}  = 0 \,. \label{U1YU1PQ4a}
	\ee}
Eq.~\eq{U1YU1PQ4} and Eq.~\eq{Modd} hint a form of the axion field ($a$) as below
\bea
a% &=& \frac{\sqrt{A}}{N_a}\left(2 \frac{PQ_{\phi} }{v_{\phi}}\, I_{\Phi} + n_u\frac{PQ_{H_u^0} }{v_{u}} \, I_{H_u^0} +  n_d \frac{PQ_{H_d^0} }{v_{d}}\, I_{H_d^0} \right) \notag\\
&=& \frac{\sqrt{A}}{N_a} \frac{2 PQ_\phi}{v_\phi} \left(  I_\phi + n_u \frac{v_\phi}{v_u} \frac{PQ_{H_u^0}}{2 PQ_\phi} I_{H_u^0} + n_d \frac{v_\phi}{v_d} \frac{2 PQ_{H_d^0}}{PQ_\phi} I_{H_d^0}\right)\,,
\label{axionnormalform}
\eea
where $A = -\lambda_5 v_u v_d v_\phi^2$ and $N_a$ is normalization factor given as
\bea
N_a = \sqrt{\left(\frac{2PQ_\phi}{v_\phi} \right)^2 + \left(n_u \frac{PQ_{H_u^0}}{v_u} \right)^2 + \left( n_d \frac{PQ_{H_d^0}}{v_d}\right)^2}\,. \label{normalizationfactor}
\eea
With $n_u = Y_{H_u^0} = -1$, $n_d = Y_{H_d^0} = 1$ and $PQ_{H_u^0} =  2 s^2_\zeta PQ_\phi$, $PQ_{H_d^0} = 2 c^2_\zeta PQ_\phi$ as shown in Tab.~\ref{tab1}, the form of axion state is rewritten as below
\bea
a &=& \frac{\sqrt{A}}{N_a} \frac{2 PQ_\phi}{v_\phi} \left(  I_\phi - \frac{v_\phi}{v_u} s^2_\zeta I_{H_u^0} +  \frac{v_\phi}{v_d} c^2_\zeta I_{H_d^0}\right)\,,
\label{axionnormalform1}
\eea
with $N_a$ is now determined as
\bea
N_a = 2 PQ_\phi \sqrt{\frac{1}{v_\phi^2} + \frac{s^4_\zeta}{v_u^2} +\frac{c^4_\zeta}{v_d^2}}\,.\label{NaCoef}
\eea
While reminding to Eq.~\eq{IHdHu}, the presentation of axion in Eq.~\eq{axionnormalform1} is represented by two imaginary components as
\bea
a &=& \frac{\sqrt{A}}{N_a} \frac{2 PQ_\phi}{v_\phi} \left[  I_\phi - v_\phi \left(\frac{s_\zeta^2}{v_u} + \frac{c_\zeta^2}{v_d} \tan \beta_1 \right) I_{H_u^0}\right]\,,
\label{axionnormalform2}
\eea
%or
%\bea
%a &=& \frac{\sqrt{A}}{N_a} \frac{2 PQ_\phi}{v_\phi} \left[  I_\phi - v_\phi \left(\frac{s_\zeta^2}{v_u} \cot \beta_1+ \frac{c_\zeta^2}{v_d}  \right) I_{H_d^0}\right]\,.
%\label{axionnormalform3}
%\eea
Substituting $\tan \beta_1$ in Eq.~\eq{oddangles} into Eqs.~\eq{axionnormalform2}, %\eq{axionnormalform3},
one has
\bea
a &=&   \frac{\sqrt{A}}{N_a} \frac{2 PQ_\phi}{v_\phi} \left(  I_\phi + \frac{v_\phi}{v_u}c_{2\zeta} I_{H_u^0} \right)\,, \label{axionnormalform4}%\\
%\mbox{or} \hs a &=& \frac{\sqrt{A}}{N_a} \frac{2 PQ_\phi}{v_\phi} \left(  I_\phi -\frac{v_\phi}{v_d}c_{2\zeta} I_{H_d^0} \right)\,, \label{axionnormalform5}
\eea
In another view point from Eq.~\eq{Oddphysicalstates}, axion physical state is expressed with three imaginary components $I_\phi, I_{H_u^0}, I_{H_d^0}$ with two mixing angle $\beta_1, \beta_2$ as
\bea
 a =  \cos \beta_2 \left( I_\phi - \sin \beta_1 \tan \beta_2 I_{H_u^0} + \cos \beta_1 \tan \beta_2 I_{H_d^0}\right) \,.\label{axionform1}
\eea
Use Eq.~\eq{IHdHu} to rewrite the form of axion state in Eq.~\eq{axionform1}, this form of axion is also presented via just two imaginary components as
\bea
a %&=& \cos \beta_2 \left( I_\phi - \sin \beta_1 \tan \beta_2 I_{H_u^0} - \cos \beta_1 \tan \beta_2 \tan \beta_1 I_{H_u^0}\right)\,,\notag\\
&=& \cos \beta_2 \left( I_\phi -2\sin \beta_1 \tan \beta_2 I_{H_u^0}\right)\,.\label{axionform2}
\eea
Equating the coefficients of Eq.~\eq{axionnormalform2} and Eq.~\eq{axionform2} yields
%\bea
%    \frac{\sqrt{A}}{ N_a}\frac{2PQ_\phi}{v_\phi} &=& \cos \beta_2\,,\notag\\
%    \frac{\sqrt{A}}{ N_a}\frac{2PQ_\phi}{v_\phi}\frac{c_{2\zeta}v_\phi}{v_u } &=& -2 \sin \beta_1 \sin \beta_2\,,  \label{zeta}
%\eea
%These equations in Eqs.~\eq{zeta} give
a constraint for the mixing of $PQ$ charges of neutral scalar components as below
\bea
\cos 2\zeta = 2 \sin 2\beta_1 \frac{v_d v_u}{v_\phi^2}\,, \label{zeta1}
\eea
Combining Eq.~\eq{zeta1} with Eq.~\eq{oddangles} leads to the constraint for $PQ$ charges of neutral components of doublet scalars via the mixing angle $\zeta$ as
\bea
c_{2\zeta} = \frac{4v_d^2 v_u^2}{v_{EW}^2 v_\phi^2}\,. \label{zeta2}
\eea
%with $v_{EW}^2 = v_d^2 + v_u^2 = 246^2 \gev^2$.
 From Eq.~\eq{zeta2}, it is shown that the mixing angle $\zeta$ depends on VEVs of scalar fields. Combining this constraint with values of $PQ$ charges in Tab.~\ref{tab1}, it is easily to see that $PQ$ charges of all particles depend on either $PQ_\phi$ or $v_\phi$ which belong to the singlet scalar causes the $PQ$ symmetry spontaneous breaking.  With the assumption $v_\phi \sim 10^{12} \gev \gg v_u,v_d \sim$ EW scale, value of $c_{2\zeta}$ in Eq.~\eq{zeta2} approximates to $10^{-21} \div 10^{-20}$. In this scenario, the mixing angle between $PQ$ charges of neutral components of doublet scalars is maximal ($\zeta \approx \frac{\pi}{4}$ rad) and $PQ$ charges of all particles in the model are represented in Tab.~\ref{tab2} as below
 \begin{table}[th]
			\resizebox{17cm}{!}{
				\begin{tabular}{|c|c|c|c|c|c|c|c|c|c|c|c|c|c|}
					\hline
			&$Q_L$		&  $\, u_R \, $ & $\, d_R\, $ & $\psi_L$ & $l_R$ & $\, \nu_L\, $ &$ H_u$ &  $H_u^0$ & $H_u^-$ & $H_d$ & $H_d^0 $ & $H_d^+$ & $\Phi$  \\ \hline
$U(1)_{PQ}$	& $\frac{PQ_\phi}{2}$ &  $-\frac{ PQ_\phi}{2}   $ & $ -\frac{ PQ_\phi}{2} $ & $ \frac{PQ_\phi}{2} $ & $ -\frac{ PQ_\phi}{2} $ & $ \frac{ PQ_\phi}{2}  $ &$  PQ_\phi  $ &  $  PQ_\phi$ & $  PQ_\phi$ & $ PQ_\phi  $ & $  PQ_\phi $ &  $ PQ_\phi $ & $ PQ_\phi$ 	 \\ \hline
		\end{tabular}}
			\caption{$U(1)_{PQ}$ charge assignments of the particle content of the model with $\zeta =\frac{\pi}{4}$ rad.}
			\label{tab2}
		\end{table}
%%%%%%%%%%%%%%%%%%%%%%%%%%%%%%%%%%%%%%%%%%%
\section{Axion - two SM fermions couplings}
As it can be seen in Eq.~\eq{YukDFSZ1}, in DFSZ-I model, at tree-level, the interactions between axion and a pair of SM fermions  are arisen from Yukawa couplings containing neutral electrical scalar
\bea
-\mathcal{L}_{Yuk}^{neutral} &=& y_u^{H_u} ( \overline{u}_L H_u^{0} u_R
+ \overline{u}_R H_u^{0*} u_L) +y_d^{H_d} ( \overline{d}_L H_d^0 d_R + \overline{d}_R  H_d^{0*} d_L)+ y_l^{H_d}(\overline{l}_L H_d^{0} l_R %+ \overline{l}_R H_d^{-} \nu_L
+ \overline{l}_R H_d^{0*} l_L )\,,
\label{yukneutral}
\eea
with axion is included in $H_u^0 =\fr{1}{\sqrt2}\left(v_u +R_{H_u^0} +i I_{H_u^0} \right)$ and  $H_d^0 =\fr{1}{\sqrt2}\left(v_d +R_{H_d^0} +i I_{H_d^0} \right)$. Combine with Eqs.~\eq{IU}, \eq{ID}, the Yukawa coupling of axion with a pair of SM fermions is given as
\bea
-\mathcal{L}_{a\bar{f}f}^{Yuk} &=& i \frac{m_u}{v_u} I_{H_u^0} \bar{u} \gamma_5 u + i \frac{m_d}{v_d} I_{H_u^0} \bar{d} \gamma_5 d + i \frac{m_l}{v_d} I_{H_u^0} \bar{l} \gamma_5 l\,,\notag\\
&\supset& -ia \sin \beta_2 \left(\frac{m_u}{v_u}\sin\beta_1 \bar{u} \gamma_5 u -\frac{m_d}{v_d} \cos \beta_1 \bar{d} \gamma_5 d - \frac{m_l}{v_d} \cos\beta_1 \bar{l} \gamma_5 l\right)\,,
\label{yukaff}
\eea
with the mass mixing angle $\sin \beta_2 \approx \tan \beta_2 \approx \beta_2 \approx 10^{-10}$ rad. Eq.~(\ref{yukaff}) shows that axion interacts with all SM fermions with tiny coupling constants. If axion is heavy enough ($m_a > 2 m_f$), it can be directly disintegrated to a pair of fermion and anti-fermion. \\
On the other hand, the particle spectrum between two type of DFSZ models is just different by an anti-doublet scalar $\tilde{H}_u$ which makes changes in Yukawa interactions but keeps the scalar potential unchanged. Hence, the coupling of axion with two SM charged leptons in DFSZ-II model is
\bea
-\mathcal{L}^{DFSZ-II}_{a\bar{l}l} = -ia \frac{m_l}{v_u}\sin \beta_2  \sin \beta_1 \bar{l} \gamma_5 l\,.\label{yukall}
\eea
These Yukawa interactions of axion and a pair of fermions in Eq.~(\ref{yukaff}) (as well as Eq.~(\ref{yukall})) help us to deal with $a \to \gamma \gamma$ decay via their contributions at one loop-level (see Fig.~\ref{agganomalous} in Sec.~\ref{AnomalousCouplings}).

\section{Anomaly couplings of axion}\label{AnomalousCouplings}
%%%%%%%%
In case that the mixing angle $\beta_2$ given in Eq.~\eq{oddangles} is tiny to neglect all terms include $\sin \beta_2$, axion contains only one component $I_\phi$ (see Eq.~\eq{Oddphysicalstates}) as
\bea
a \approx I_\phi \cos \beta_2\,. \label{axion3}
\eea
It is well-known that this axion generates an anomalous coupling from the effective Lagrangian
\bea
\mathcal{L}_a  \supset \fr{a}{f_a} \fr{g_s^2 \, }{32 \pi^2} G \tilde{G} +
	\fr{1}{4} g_{a\, \ga\ga}  a F \tilde{F} + \fr{\pa^\mu a}{v_a} J_\mu^{PQ}\,,
\eea
with
	\be g_{a\, \ga\ga} = \fr{\al}{2 \, \pi  \, f_a}\, \fr E N \,.
	\label{anomalouscoupling}
	\ee
In the model under consideration, the $QCD$ anomalous coefficient is defined as ~\cite{luzio}
\be
\label{nov4}
N=\sum_\mathcal{Q} N_\mathcal{Q} = \sum_\mathcal{Q} PQ({\mathcal{Q})}\,
 n_C (\mathcal{Q})\,  n_I(\mathcal{Q})\,  T(\cal{C}_{\mathcal{Q}}) \,,
\ee
in which,
	\begin{enumerate}
		\item $ n_C (\mathcal{Q})$ and $n_I(\mathcal{Q})$ denote the dimension of the  colour and weak isospin representations, respectively, for example:  ($n( {\bf  T) }= 3$ for triplet and $n ({\bf S})	= 1$ for singlet).
		\item $T(\cal{C}_\mathcal{Q})$ is the colour Dynkin index, for example
		$T(3)=1/2$.
	\end{enumerate}

\begin{figure}[h]
\begin{center}
    \includegraphics[width=8cm]{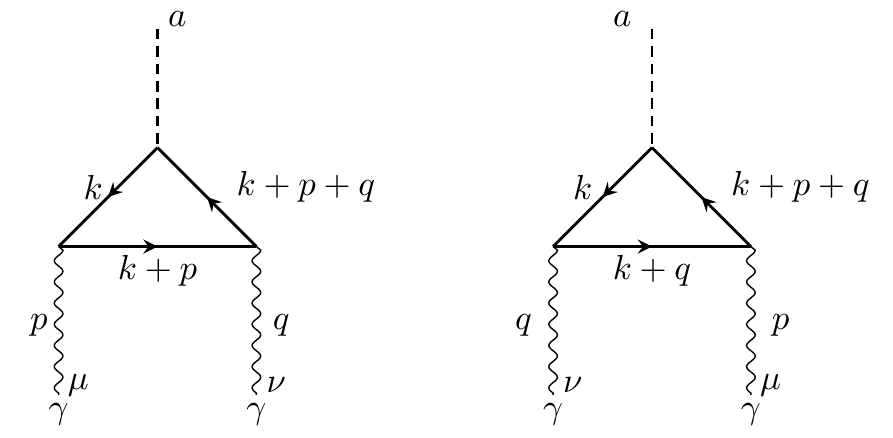}
\end{center}
\caption{Feynmann diagrams for anomaly coupling of axion-photon-photon with a fermion-loop, all momenta are coming in.}% at one-loop level. 
\label{agganomalous}
\end{figure}

%Taking into account  $PQ q_L = - PQ q_R$
Use Eq.~(\ref{a161}) and values of $PQ$ charges in Tab.~\ref{tab1}, the $QCD$ anomalous coefficient is read as
\bea
 N_{(DFSZ-I)} & = & n_c     \fr 1 2 [ (PQ_{ Q_{ a L}}  - PQ_{ u_{a R})} +
 (PQ_{ Q_{ a L}}  - PQ_{ d_{a R})}] \crn
 & = & \fr 3 2 ( PQ_{ H_{u}} +  PQ_ {H_{d}}) %= \vhb{3 PQ_\phi}
 \,.
\label{Ndfsz1}
\eea
Now we turn to  electromagnetic $[U(1)_Q]^2\times U(1)_{PQ}$ anomalous coefficient
\be
E  = \sum\limits_{i= charged}\left( PQ_{f_{iL}} - PQ_{f_{i R}}\right)(Q_{f_i})^2\,.
\label{EU1QU1PQ}
	\ee	
Taking $PQ$ charges from Table \ref{tab1} and the color number of quarks  $n_c = 3$, one has
\bea
E_{DFSZ-I} & = &   n_c 3 [(PQ_{ u_{ a L}}  - PQ_{ u_{a R}}) ( \fr 2 3 )^2
+ (PQ_{ d_{ a L}}  - PQ_{ d_{a R}}) ( -\fr 1 3 )^2] +  3  (PQ_{ l_{ a L}}  - PQ_{ l_{a R}}) (-1)^2\crn
& = &   n_c 3 [PQ_{H_u} ( \fr 2 3 )^2
+ PQ_{H_d} ( -\fr 1 3 )^2] +  3  (PQ_{H_d} (-1)^2\crn
& = & 4 PQ_{H_u} +  PQ_{H_d} + 3  PQ_{H_d} \crn
& = & 4 (PQ_{H_u} +  PQ_{H_d}) %= \vhb{8 PQ_{\phi}}
\,.
\label{Edfsz1}
\eea
Reminding to Eq.~(\ref{PQs11}), values of  $N$ (in Eq.~(\ref{Ndfsz1})) and $E$ (in Eq.~(\ref{Edfsz1})) are turned out to be independent from $PQ$ charges of fermions. Their values just depend on $PQ$ charge of singlet scalar which causes the spontaneous breaking of $U(1)_{PQ}$ symmetry.
\bea  N_{(DFSZ-I)}= 3 PQ_\phi \,, \hs E_{DFSZ-I} = 8 PQ_{\phi}\,. \label{NEdfsz1}
\eea
%Combine Eq.~\eq{Ndfsz1} and Eq.~\eq{Edfsz1},
From Eq.~(\ref{NEdfsz1}), the ratio of these two above coefficients is a constant as
%Then, in the model under consideration, one gets
\be
		\frac{E_{DFSZ-I}}{N_{DFSZ-I}}  = \frac{8}{3}\,. \label{ENratio}
	\ee
    This ratio in Eq.~(\ref{ENratio}) is a constant which is even independent from $PQ$ charge of the singlet scalar $\Phi$.
Performing current algebra, one gets the axion mass given by \cite{sr}
\be
		g_{a\ga} =\fr{\al}{2 \pi f_a} \left(\fr E N - \fr{2}{3}\fr{4+\frac{m_u}{m_d} + \fr{m_u}{m_s} }{1+\fr{m_u}{m_d}  +\fr{m_u}{m_s}}\right) = \fr{\al}{2 \pi f_a} \left(\fr E N  - 1.92 \right) \,.
		\label{gagaDFSZ1}
	\ee
Substituting Eq.~\eq{ENratio} into Eq.~\eq{gagaDFSZ1}, the axion-photon coupling is given as follows
\be
		g_{a\ga}^{DFSZ-I}  = \fr{\al}{2 \pi f_a} \left( \fr 8 3  - 1.92 \right) \,.
		\label{nov1815}
	\ee
This result is consistent with that in Ref.~\cite{julio2} and get the larger absolute value than itself in the  DFSZ-II model ~\cite{DFSZ1,DFSZ2} where $\frac{E}{N} = + \frac{2}{3}$ (see Appendix \ref{AppA}). The reason makes the ratio $\frac{E}{N}$ of $DFSZ - II$ model to be different from itself in $DFSZ - I$ model is because of the different Yukawa interaction term of leptons. This term makes $PQ$ charges of particles in $DFSZ - II$ model changing in comparison with themselves in $DFSZ - I$ model.\\
It is worth noting that this anomaly coupling of axion and two photons is forbidden at tree-level but it can happen at one-loop level with fermion-loop as shown in Fig.~\ref{agganomalous}.

\section{Axion - neutral gauge bosons  couplings}
In our study, the mixtures of scalar fields are considered then the physical state of $DFSZ$ axion contains three components as in Eq.~(\ref{axionform1}). It is recognized that there are not only anomalous couplings of axion into a pair of photons as shown in Sec.\ref{AnomalousCouplings} but also triple-couplings of axion-photon-photon arisen from the kinetic term of scalar fields. This happens because of the spontaneous breaking of $U(1)_{PQ}$ global symmetry and the spontaneous breaking of $SU(2)_L \times U(1)_Y$ local symmetries are studied together. Because of considering the components that either trigger the breaking of local symmetries or being included in physical state of axion, a new kind of interaction between axion and two neutral gauge bosons at tree-level are arisen with tiny couplings depending on tiny mass mixing angles defined from diagonalizing $CP$-odd mass mixing matrix.\\ %This extremely small coupling explains why it is very difficult to search for such a new kind of interaction between axion and two neutral gauge bosons at tree-level.\\
%First of all,
It shoud be started by expanding the scalar around its VEV as
\be \phi^\prime = \langle \phi^\prime \rangle + \phi\,.
\label{o221}
\ee
The interaction of axion and two photons is arisen from the below kinetic term of scalar fields $(\phi=\Phi, H_u, H_d)$
\bea
\left(D^\mu \phi^\prime \right)^\dag D_\mu \phi & = & \pa^\mu \phi \pa_\mu \phi + i \left[\, \langle \phi^\prime \rangle^\dag P_\mu^{\phi} \pa^\mu  \phi - \pa_\mu \phi^\dag P^{\phi, \mu}  \langle \phi^\prime \rangle \right]\crn
&+& i \left[\, \phi^\dag P^\phi_\mu \pa^\mu \phi - \pa_\mu \phi^\dag  P^{\phi, \mu}\phi
 \right] + \phi^{'\, \dag}  P_\mu^{\phi}  P^{\phi, \mu}\phi^\prime\,,
\label{ScalarKinertic}
\eea
  The last term in {Eq.~\eq{ScalarKinertic} is expressed via charged and neutral currents in $P_\mu^\phi$ as}
\bea
 \phi^{\prime\, \dag}  P_\mu^{\phi}  P^{\phi, \mu}\phi^\prime & = & \langle \phi^\prime \rangle^\dag   P_\mu^{\phi}  P^{\phi, \mu}  \langle \phi^\prime \rangle  + \phi^{, \dag}  P_\mu^{\phi}  P^{\phi, \mu}\phi
\crn
&+& i \left[ \langle \phi^\prime \rangle^\dag   P_\mu^{\phi}  P^{\phi, \mu}\phi -  \phi^\dag  P_\mu^{\phi}  P^{\phi, \mu}  \langle \phi^\prime \rangle \right]\,.
\label{3rdSK}
\eea
In Eq.~\eq{3rdSK}, the first term generates masses for gauge bosons, the second one gives couplings of two gauge bosons with two scalar fields and the last one provides coupling of two gauge bosons and one scalar field. This means all couplings of two gauge bosons and one axion are able to be studied in detail. 

\begin{figure}[h!]
	\centering
	\begin{tabular}{c}
		\includegraphics[width=6cm]{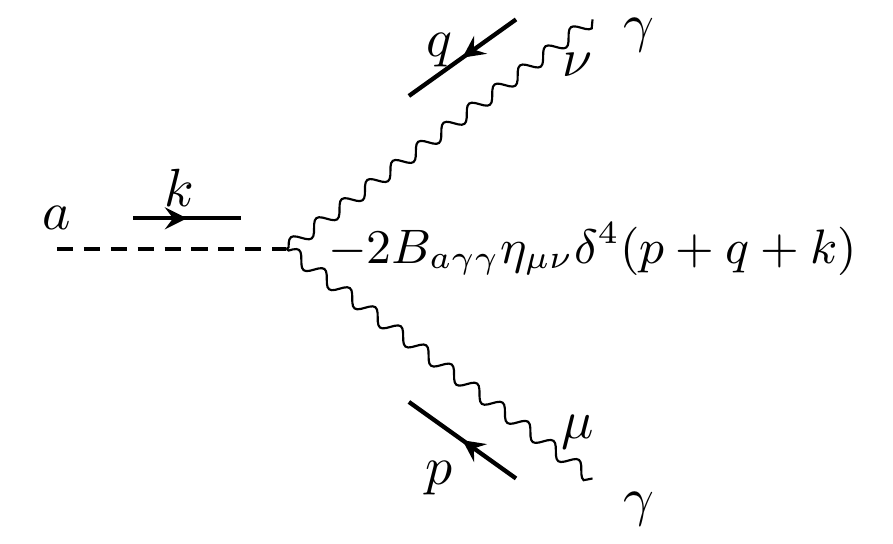}
	\end{tabular}%
	\caption{Feynman rules for triple-coupling of axion-photon-photon with all momenta are coming in.}
	\label{fig_agg}
\end{figure}

In this section, we focus on the last term containing the interaction of axion with neutral gauge bosons
\be
 \mathcal{L}_{ a } ^{NGB}\in \sum_{\phi=H_u, H_d}
 i \left[ \langle \phi^\prime \rangle^\dag   P_\mu^{\phi}  P^{\phi \mu}\phi -  \phi^\dag  P_\mu^{\phi}  P^{\phi \mu}  \langle \phi^\prime \rangle \right]\,,
\label{LaNgNg}
\ee
with %$\phi = H_u, H_d$ while
$P_\mu$ given in Eq.~\eq{Pmu}, gauge states are presented via physical state $A_\mu, Z_\mu$ (see Eq.~\eq{PhysNG}) and imaginary components of scalars are presented as in Eqs.~\eq{IU}, \eq{ID}. Consequently, there are some available triple-couplings of axion and two gauge boson at tree-level as below
\bea
\mathcal{L}( a \gamma \gamma )& = & -i %\frac{-g^2 v_d}{2} \frac{\cos^2 2\theta_W}{\cos^2 \theta_W} \frac{\sin \beta_2}{\cos \beta_1} a \gamma \gamma \equiv -i C_{a\gamma\gamma} a \gamma \gamma\,, \label{Lagg} \\ %
\frac{g^2}{4} \frac{\cos^22\theta_W}{\cos^2 \theta_W} \sin \beta_2 \left( v_u \sin \beta_1 -v_d \cos \beta_1\right) a \gamma \gamma \equiv -i B_{a\gamma\gamma} a \gamma \gamma\,, \label{Lagg} \\
\mathcal{L}( a ZZ )& = & -ig^2 \sin \beta_2\sin^2\theta_W \left(v_u \sin \beta_1 - v_d \cos \beta_1 \right) aZ_\mu Z^\mu \equiv -i B_{aZZ}Z Z a \,, \label{LaZZ}\\
\mathcal{L}( a \gamma Z )& = & -i g^2 \cos 2\theta_W \tan \theta_W \sin \beta_2 \left( v_u \sin \beta_1 - v_d \cos \beta_1\right)a\gamma Z \equiv -i B_{a\gamma Z}a\gamma Z\,. \label{LagZ}
\eea
The coupling of axion and two photons in Eq.~(\ref{Lagg}) is presented in Fig.~\ref{fig_agg} with
\bea
B_{a\gamma\gamma} = \frac{g^2 }{4} \frac{\cos^2 2\theta_W\sin \beta_2}{\cos^2 \theta_W} \left(v_u \sin \beta_1 - v_d \cos \beta_1 \right) %\notag\\
%&=& \frac{-g^2 v_d}{2} \frac{\cos^2 2\theta_W}{\cos^2 \theta_W} \frac{\sin \beta_2}{\cos \beta_1}
\,. \label{Cag}
\eea
This triple-coupling of axion-photon-photon is different from coupling of Primakoff effect \cite{s5,agu,s6,s7} where axion interacts with photon and dark-photon.

\section{Conclusions}

Because of the mixture between scalars in the model under consideration, form of axion field must exactly include three components coming from the imaginary parts of three scalar fields. This result is totally different from the thought of axion with only one imaginary component of singlet scalar. By requiring all terms of Yukawa interaction of the model must be preserved under $U(1)_{PQ}$ transformation,  $PQ$ charges of all particles are predicted. It is consistently that these values of $PQ$ charges are just depended on two parameters that belong to the singlet scalar causing the spontaneous breaking of $U(1)_{PQ}$ symmetry. Those two parameters are $PQ$ charge of the singlet scalar and the mixing angle between $PQ$ charges of two neutral electrical components of doublet scalars. The constraint for this tiny mixing angle is also pointed out. By using defined $PQ$ charges of scalars and requirement of invariant under $U(1)_Y$ and $U(1)_Y \times U(1)_{PQ}$, the general form of axion in the model is also predicted. And it is interesting that this general form of axion is consistent with physical state of axion determined from the mass mixing matrix of $CP$-odd sector. With a structure of three imaginary components of scalars, axion interacts either with a pair of SM fermions or a pair of neutral gauge bosons at tree-level.  These interactions are arisen from Yukawa interactions and kinetic terms of scalar fields, respectively.  Especially, the triple-coupling of axion-photon-photon at tree-level is really needed to be studied as carefully as possible cause it can help for experiments that are searching for axion. Other triple-coupling such as $aZ\gamma$ is also need to be paid attention when studying on axion in any Beyond Standard Models, too.

\section*{Acknowledgements}
This research has received funding from  National Foundation for Science and Technology Development (NAFOSTED) under grant number 103.01-2023.45.

%\section*{Conflict of Interest} The authors declare that they have no %conflicts of interest.


\begin{thebibliography}{99}

\bibitem{DFSZ1} A. Zhitnitsky, Sov. J. Nucl. Phys. {\bf 31} (1980) 260.

\bibitem{DFSZ2} M. Dine, W. Fischler, M. Srednicki, Phys. Lett. B {\bf 104} (1981) 199-202.
\bibitem{vphi1} J. Preskill, M. B. Wise and F.Wilczek, Phys. Lett. B120, 127 (1983);

\bibitem{vphi2} L. F. Abbott and P. Sikivie, Phys. Lett. B120, 133 (1983);
\bibitem{r1} Moslem Ahmadvand, Fazlollah Hajkarim,
%Lepton g−2 and W-boson mass anomalies in the DFSZ axion model,
 Eur. Phys. J. C (2023) 83: 1021, arXiv:2302.09610 [hep-ph]
%DOI: 10.1140/epjc/s10052-023-12195-2

\bibitem{r2} H. N. Long, H. T. Hung, V. H. Binh, A. B. Arbuzov:  Peccei-Quinn charges in the 3-3-1 model with $U(1)_{B-L}$ symmetry, arXiv:2412.04269  [hep-ph]

\bibitem{julio2} A. G. Dias, J. Leite, José W. F. Valle, C. A. Vaquera-Araujo, Phys. Let.  B 810 (2020) 135829, arXiv:2008.10650 [hep-ph].

\bibitem{luzio} L. Di Luzio, M.  Giannotti, E. Nardi and L. Visinelli,  Phys. Rept. 870 (2020) 1,  arXiv: 2003.01100 [hep-ph]

\bibitem{longpal} H. N. Long and P. B. Pal,
%: Nucleon instability in a
%supersymmetric  $\mbox{SU(3)}_C\otimes \mbox{SU(3)}_L\otimes
%\mbox{U(1)}$ model.,  ICTP, Trieste preprint IC/97/202,
 {\it Mod. Phys. Lett.} {\bf A13}, (1998) 2355
- 2360,  [arXiv: hep-ph/9711455]


 \bibitem{sr} M. Srednicki, Nucl. Phys. B 260 (1985) 689.

 \bibitem{s5} P. Sikivie, Phys. Rev. Lett. 51, 1415 (1983); Phys. Rev. D32, 2988 (1985).
		
\bibitem{agu} I. G. Irastorza, J. Redondo, Progress in Particle and Nuclear Physics, 102  (2018)  89, 	arXiv:1801.08127.
		
		
\bibitem{s6} K. Van Bibber, N. R. Dagdeviren, S. E. Koonin, A. K. Kerman, and H. N. Nelson,
		Phys. Rev. Lett. 59, 759 (1987).
		
\bibitem{s7}H. N. Long, D. V. Soa, and T. A. Tran,
		%Electromagnetic detection of axions. [arXiv: hep-ph/9507392],
		{\it Phys. Lett.} {\bf B357},
		(1995) 469, [arXiv: hep-ph/9507392]
\end{thebibliography}
\end{document}